\documentstyle[12pt]{article}
\def\lsim{\mathrel{\lower2.5pt\vbox{\lineskip=0pt\baselineskip=0pt
           \hbox{$<$}\hbox{$\sim$}}}}
\def\gsim{\mathrel{\lower2.5pt\vbox{\lineskip=0pt\baselineskip=0pt
           \hbox{$>$}\hbox{$\sim$}}}}

\makeatletter

\@addtoreset{equation}{section}
\makeatother

\begin{document}
\setlength{\baselineskip}{8mm}
\begin{titlepage}
\begin{flushright}
\begin{tabular}{c c}
& {\normalsize  hep-ph/9703250} \\
& {\normalsize  DPNU-97-14} \\
& {\normalsize March 1997}
\end{tabular}
\end{flushright}
\vspace{5mm}
\begin{center}
{\large \bf Supersymmetry Breaking Without A Messenger Sector}\\
\vspace{15mm} 
Naoyuki Haba\footnote{E-mail:\ haba@eken.phys.nagoya-u.ac.jp, 
Address after April 1, 1997: Faculty of Engineering, 
Mie University, Mie, JAPAN 514}, 
Nobuhito Maru\footnote{E-mail:\ maru@eken.phys.nagoya-u.ac.jp},   
and 
Takeo Matsuoka\footnote{E-mail:\ matsuoka@eken.phys.nagoya-u.ac.jp} \\ 

{\it 
Department of Physics, Nagoya University \\
           Nagoya, JAPAN 464-01 \\
}
\end{center}

\vspace{10mm}

%%%%%%%%%%%%%%%%%%%%%%%%%%%%%%%%%%%%%%%%%%%%%%%%%%%%%%%%%%%%%%%%%%%
%%%%%%%%%%%%%%%%%%%    ABSTRACT    %%%%%%%%%%%%%%%%%%%%%%%%%%%%%%%%
%%%%%%%%%%%%%%%%%%%%%%%%%%%%%%%%%%%%%%%%%%%%%%%%%%%%%%%%%%%%%%%%%%%

\begin{abstract}

We propose a simple gauge-mediated supersymmetry 
breaking model in which suitable soft breaking 
masses are dynamically generated without relying 
on a messenger sector. 
This model is constructed as an extention 
of the 3\,-2 model and needs no fine-tuning of 
parameters. 
The dynamical supersymmetry breaking sector 
contains non-renormalizable interactions 
and vector-like fields. 
Non-renormalizable terms are characterized 
by the couplings of $O(1)$ in units of 
the Planck scale. 
Hierarchy of various mass scales in the model 
arises from the non-renormalizable terms. 
The vacuum of this model conserves 
the color and electro-weak symmetry. 
We also discuss a possible solution to 
the $\mu$ problem in this framework. 

\end{abstract}

%\vspace{2cm}
%\begin{flushleft}
%PACS number(s): 12.60.Jv, 11.30.Qc, 14.80.Ly
%\end{flushleft}

\end{titlepage}

%%%%%%%%%%%%%%%%%%%%%%%%%%%%%%%%%%%%%%%%%%%%%%%%%%%%%%%%%%%%%%
%%%%%%%%%%%%%%%%%%%%%%%  Section 1  %%%%%%%%%%%%%%%%%%%%%%%%%%
%%%%%%%%%%%%%%%%%%%%%%%%%%%%%%%%%%%%%%%%%%%%%%%%%%%%%%%%%%%%%%

\section{Introduction}

The dynamical supersymmetry breaking (DSB) 
provides an attractive explanation of the large 
hierarchy between the Planck scale and 
the electro-weak scale. 
Many models which exhibit the DSB have 
been proposed \cite{DSB1,DSB3,DSB4,IT}. 
Since these models by themselves have nothing 
to do with our visible world, 
we need some mechanisms of mediating 
supersymmetry breaking to the low-energy 
observable sector. 
For phenomenological implications it is of 
great importance to clarify how the supersymmetry 
breaking is communicated to the visible sector. 
Two ways of transmitting supersymmetry breaking 
to the visible sector have been proposed. 
The first is that gravitational interactions in 
$N=1$ supergravity theory play its role. 
The second is that gauge interactions 
play a role of the messenger. 
As for the latter one, the scenario given 
in Ref. \cite{DSB1} suffers from a problem 
that the QCD coupling becomes strong a few decades 
above the scale of the supersymmetry breaking. 
This difficulty originates from the situation 
of the model in which the standard model gauge group 
$G_{st}\equiv SU(3)_c \times SU(2)_L \times U(1)_Y$ 
is taken as the global symmetry of the DSB sector. 
In the model given in Refs. \cite{456}, 
this difficulty was avoided by introducing 
the ``messenger sector'' which connects 
the DSB sector with the standard model. 
In the messenger sector scenario the supersymmetry 
is broken at the energy as low as $O(100\,{\rm TeV})$. 
In such a model several attractive features 
are derived. 
Namely, we obtain enough degeneracy among squarks 
and sleptons to ensure adequate suppression of 
flavor changing neutral currents (FCNC). 
Further, no new $CP$ phases are induced in soft 
supersymmetry breaking parameters. 
This implies the absence of large particle 
electric dipole moment (EDM).

Recently, however, it has been pointed out that 
in the original model of the messenger sector scenario 
QCD color symmetry is not conserved \cite{RAN}. 
In Ref. \cite{456}, 
the superpotential in the messenger sector 
is assumed to be 
%%%%%%%%%%%%%%%%%%%%%%%%%%%%%%%%%%%%%%%%%%%
\begin{equation}
\label{mess}
W_{{\rm mess}} = k_1 \phi^+ \phi^- S + 
\frac{1}{3} \lambda S^3 
+ k_3 S q\bar{q} + k_4 S l\bar{l}\,, 
\end{equation}
%%%%%%%%%%%%  mess  %%%%%%%%%%%%%%%%%%%%%%%
where the vector-like multiplets 
$q, \bar{q}$ transform 
as ${\bf 3}$, ${\bf 3^*}$ 
under color $SU(3)_C$ and $l, \bar{l}$ 
transform as ${\bf 2}$ under $SU(2)_L$. 
$S$ is a singlet 
and $\phi^{\pm}$ have $\pm1$ charges of the messenger 
gauge group $U(1)_m$. 
Since the soft supersymmetry breaking masses 
squared for $\phi^+$ and $\phi^-$ become negative, 
$\phi^+$ and $\phi^-$ develop non-zero vacuum expectation 
values (VEVs). 
If the minimization of the scalar potential yields 
non-vanishing $\langle S \rangle$ and 
$\langle F_S \rangle$, 
the supersymmetry breaking is transmitted 
to the singlet $S$ by the interaction 
$\phi^+ \phi^- S$ in Eq. (\ref{mess}). 
This is a crucial point for communicating 
supersymmetry breaking to the visible sector 
since gauginos, squarks and sleptons masses 
are radiatively induced by the interactions 
$S q\bar{q}$ and $S l\bar{l}$. 
However, in the true vacuum $\langle S \rangle$ and 
$\langle F_S \rangle$ vanish and at the same time 
QCD color is violated \cite{RAN}.

Many attempts have been made in order to resolve 
the problem \cite{Many,ours}. 
In the previous paper \cite{ours}, 
the authors proposed a simple model in which 
the messenger gauge interaction is not necessarily 
needed and the vacuum does not break the color 
and electro-weak symmetry. 
So the model is free from the difficulty encountered 
in the original messenger sector scenario. 
In the model the effective theory from the DSB 
sector serves as the messenger sector 
and non-vanishing VEVs of $S$ and $F_S$ are ensured 
by the dynamics in the DSB sector. 
Thus, supersymmetry breaking is transmitted from 
DSB sector to the visible sector through 
the interactions $S q\bar{q}$ and $S l\bar{l}$. 
No elaborate messenger sector is needed to
achieve the feed-down of the supersymmetry breaking. 
The strategy of this paper is to construct the model 
which satisfy the following requirements:
%%%%%%%%%%%%%%%%%%%%%%%%%%%%%%%%%%%%%%%%%%%%%%
\begin{enumerate}
  \item No messenger gauge interactions. 
  \item Communication of supersymmetry breaking to 
   the visible sector without violating color symmetry. 
  \item The soft supersymmetry breaking masses coming 
   predominantly from the standard model gauge 
   interaction rather than the gravitational interaction. 
  \item No fine-tuning of parameters or very small 
   coupling constants. 
  \item No $\mu$ problem. 
\end{enumerate}
%%%%%%%%%%%%%%%%%%%%%%%%%%%%%%%%%%%%%%%%%%%%%%
Although the model proposed in the previous paper 
\cite{ours} satisfies the first three among 
the above requirements, 
the model relies on an extremely small 
coupling constant and a solution to the 
$\mu$ problem is not presented.

In this paper we propose a renewed model 
which is in line with the above five requirements. 
The superpotential in the present model consists 
of non-renormalizable interactions. 
Since the superpotential involves a scale explicitly, 
the present model, in a sense, seems to break
the philosophy in Ref. \cite{456} that 
all masses should arise through dimensional 
transmutation. 
However, in the case that the suppression factor 
of non-renormalizable terms is the Planck scale, 
the model does not contradict with the philosophy. 
This is because there exists the supergravity for 
the underlying theory. 
If we do not take the supergravity into account, 
models of the DSB suffer from a massless Goldstone 
fermion and a massless $R$-axion. 
A massless Goldstone fermion, which appears as 
a result of the spontaneous supersymmetry breaking, 
is absorbed by the gravitino. 
An $R$-axion can gain a mass by breaking the $R$ 
symmetry explicitly to cancel the cosmological 
constant \cite{BAG}. 
Although soft masses are induced by both 
gravitational interactions and gauge interactions, 
the contribution of gauge interactions is 
dominant in the present model. 
The supersymmetry breaking scale is the intermediate 
scale of order $10^{10}\,{\rm GeV}$.

This paper is organized as follows. 
After explaining the model proposed in the previous 
paper \cite{ours} in Section 2, 
we proceed in Section 3 to exhibit a renewed model. 
A possible solution to the $\mu$ problem is 
discussed in Section 4. 
Section 5 is devoted to summary.

%%%%%%%%%%%%%%%%%%%%%%%%%%%%%%%%%%%%%%%%%%%%%%%%%%%%%%%%%%%%%%%
%%%%%%%%%%%%%%%%%%%%%%  Section 2  %%%%%%%%%%%%%%%%%%%%%%%%%%%%
%%%%%%%%%%%%%%%%%%%%%%%%%%%%%%%%%%%%%%%%%%%%%%%%%%%%%%%%%%%%%%%

\section{Fine-tuning in the DSB models}

To begin with, let us briefly review the parts 
of the 3\,-2 model which are relevant to our study. 
The 3\,-2 model is the simplest calculable model 
with the dynamical supersymmetry breaking \cite{DSB1}. 
The gauge group of this model is 
$SU(3) \times SU(2)$. 
This model contains chiral superfields 
%%%%%%%%%%%%%%%%%%%%%%%%%%%%%%%%%%%%%%%%%%%%%%%%%%%%%%
\begin{equation}
\label{DSBfield}
 Q \; ({\bf 3,2})_{1/3}, \;\; 
 \overline{U}\; ({\bf 3^*,1})_{-4/3}, \;\; 
 \overline{D}\; ({\bf 3^*,1})_{2/3}, \;\;
 L \; ({\bf 1,2})_{-1}\;, 
\end{equation}
%%%%%%%%%%%%%  DSBfield  %%%%%%%%%%%%%%%%%%%%%%%%%%%%%
where the subscript denotes the quantum number 
of a global $U(1)$ symmetry of 
the model
\footnote{
In the ordinary ``messenger sector'' scenario 
this $U(1)$ symmetry is localized to mediate 
the supersymmetry breaking \cite{456}. 
The present model, however, does not need 
to localize it.
}. 
The superpotential of the 3\,-2 model is given by 
%%%%%%%%%%%%%%%%%%%%%%%%%%%%%%%%%%%%%%%%%%%%%%%%%%%
\begin{equation}
\label{W32}
 W_{3-2}  = \lambda _1X_1 + {\Lambda^7 \over X_3}, 
\end{equation}
%%%%%%%%  W32  %%%%%%%%%%%%%%%%%%%%%%%%%%%%%%%%%%%%
where $X_i$'s are gauge invariant combinations 
%%%%%%%%%%%%%%%%%%%%%%%%%%%%%%%%%%%%%%%%%%
\begin{equation}
\label{Xi}
 X_1 \equiv Q \overline{D} L \; , \;\; 
 X_2 \equiv Q \overline{U} L \; , \;\; 
 X_3 \equiv {\rm det}Q \overline{Q} \; . \;\; 
\end{equation}
%%%%%%%  Xi  %%%%%%%%%%%%%%%%%%%%%%%%%%%%%
Here we use the notation $Q \equiv ({U},{D})$ and 
$\overline{Q} \equiv ( \overline{U}, \overline{D})$. 
The first term of $W_{3-2}$ is the most general 
renormalizable superpotential which is invariant 
under the symmetries. 
A coupling constant $\lambda _1$ is assumed to be 
very small so as $\lambda _1 ^{1/7} \ll 1$. 
The second term in Eq. (\ref{W32}) is dynamically 
generated by the one-instanton effect and yields 
the runaway behavior of the scalar potential. 
This runaway behavior is stabilized due to 
the lifting effect of the first term. 
The smallness of $\lambda_1$ ensures that 
the superfields develop large VEVs 
while maintaining only a small vacuum energy. 
Relying on a very small coupling is shared 
in common with a number of models which 
invoke the runaway-behaving potential. 
Now we consider the situation that 
the $SU(3)$ coupling is larger than the $SU(2)$ 
coupling. 
$\Lambda$ denotes the scale at which the $SU(3)$ 
coupling becomes large
\footnote{
If we consider the case where $SU(2)$ 
dynamics is dominant compared to $SU(3)$ dynamics, 
supersymmetry is broken due to the quantum deformation 
of the moduli space \cite{IT}.
}. 
{}From the superpotential (\ref{W32}), 
VEVs of elementary fields $Q, L, ..$ are estimated 
to be of order $\Lambda / \lambda_1^{1/7}( \equiv v)$. 
Since we have the relation $v \gg \Lambda$, 
gauge couplings are effectively weak at the scale $v$. 
Therefore, effective fields $X_i$'s can be regarded as 
the product of elementary fields and 
K\"ahler potential has the canonical form for 
elementary fields. 
As $F$-terms of elementary fields turn out to be of 
$O( \lambda_1 v^2)$, 
the supersymmetry breaking occurs in the 3\,-2 model. 
After integrating out the heavy degrees of freedom 
at the scale $\Lambda$, 
we can construct the effective theory below 
the scale $\Lambda$ which is described in terms 
of the light fields $X_i$'s. 
The K\"ahler potential in the effective theory 
is given by \cite{DSB1,BAG} 
%%%%%%%%%%%%%%%%%%%%%%%%%%%%%%%%%%%%%%%%%%%%
\begin{equation}
\label{K}
   K = 3 \left( t + \frac {B}{t} \right), 
\end{equation}
%%%%%%%%%  K  %%%%%%%%%%%%%%%%%%%%%%%%%%%%%%
where 
%%%%%%%%%%%%%%%%%%%%%%%%%%%%%%%%%%%%%%%%%%%%%%%%%%%%%
\begin{eqnarray}
   t & \equiv & (A + \sqrt{A^2-B^3})^{1/3}
               + (A - \sqrt{A^2-B^3})^{1/3}, \nonumber  \\
   A & \equiv & {1 \over 2}
               (X_1^{\dagger} X_1 + X_2^{\dagger} X_2),  \\
   B & \equiv & {1 \over 3}
                     \sqrt{X_3^{\dagger} X_3}. \nonumber 
\end{eqnarray}
%%%%%%%%%%%%%%%%%%%%%%%%%%%%%%%%%%%%%%%%%%%%%%%%%%%%%

In the previous paper \cite{ours} it is assumed 
that the superpotential is of the form 
%%%%%%%%%%%%%%%%%%%%%%%%%%%%%%%%%%%%%%%%%%%%%%%%%%%%
\begin{equation}
\label{W321}
    W = W_{3-2} + W_{m} + \frac {\lambda_2'}{M^5}
                   X_2 \overline{E} X_3 
\end{equation}
%%%%%%%%%%  W321  %%%%%%%%%%%%%%%%%%%%%%%%%%%%%%%%%%
with 
%%%%%%%%%%%%%%%%%%%%%%%%%%%%%%%%%%%%%%%%%%%%
\begin{equation}
\label{Wm}
    W_{m} = \frac {h_1}{M^2} X_1 q \overline{q} + 
            \frac {h_2}{M^2} X_1 l \overline{l}\; . 
\end{equation}
%%%%%%%%%%  Wm  %%%%%%%%%%%%%%%%%%%%%%%%%%%%
Here $M$ is set equal to the reduced Planck scale 
$M_{{\rm Planck}}$. 
Vector-like fields $q$, $\overline{q}$ 
and $l$, $\overline{l}$ are singlets under 
the 3\,-2 model gauge group $SU(3) \times SU(2)$ 
but charged under the standard model gauge group 
$G_{st} = SU(3)_c \times SU(2)_L \times U(1)_Y$ as 
%%%%%%%%%%%%%%%%%%%%%%%%%%%%%%%%%%%%%%%%%%%%%%%%%
\begin{equation}
 q \; ({\bf 3, 1}, -2/3), \;\; 
 \overline{q}\; ({\bf 3^*,1}, 2/3), \;\;
 l \; ({\bf 1,2}, 1),     \;\; 
 \overline{l}\; ({\bf 2,1}, -1) \;. \;\;
\end{equation}
%%%%%%%%%%%%%%%%%%%%%%%%%%%%%%%%%%%%%%%%%%%%%%%%%
Note that the global $U(1)$ symmetry offered 
in Eq. (\ref{DSBfield}) is gauged 
and the singlet superfield $\overline{E}$ is 
included to cancel the gauged $U(1)$ anomaly. 
Further, a non-anomalous $U(1)_R$ symmetry is 
introduced. 
However, the $U(1)$ gauge interaction and 
the last term in Eq. (\ref{W321}) do not play 
an essential role in communicating the supersymmetry 
breaking. 
The fields $X_1$ and $X_3$ correspond to 
the singlet field $S$ in Ref. \cite{456}. 
The above superpotential yields non-vanishing VEVs 
for $X_1$, $X_3$ and $F_{X_1}$, $F_{X_3}$ 
in the true vacuum, which are 
%%%%%%%%%%%%%%%%%%%%%%%%%%%%%%%%%%%%%%%%%%
\begin{eqnarray}
   \langle X_1 \rangle & = & O(v^3), \qquad 
   \langle F_{X_1} \rangle = O(\lambda _1 v^4), \\
   \langle X_3 \rangle & = & O(v^4), \qquad 
   \langle F_{X_3} \rangle = O(\lambda _1 v^5). 
\end{eqnarray}
%%%%%%%%%%%%%%%%%%%%%%%%%%%%%%%%%%%%%%%%%%
Confronting the result with phenomenological 
constraints on gaugino, squark and slepton masses, 
the authors have obtained 
\cite{ours} 
%%%%%%%%%%%%%%%%%%%%%%%%%%%%%%%%%%%%%%%%%%%%%%%%%%
\begin{eqnarray}
     10^{14}\,{\rm GeV} < & v & < 10^{16}\,{\rm GeV}, \\
     10^{-11} < & \lambda _1 & < 10^{-9}. 
\end{eqnarray}
%%%%%%%%%%%%%%%%%%%%%%%%%%%%%%%%%%%%%%%%%%%%%%%%%%
This means that the model needs an unnatural 
fine-tuning of the parameter $\lambda_1$. 
In order to find a phenomenologically viable model 
it is necessary to resolve this problem.

%%%%%%%%%%%%%%%%%%%%%%%%%%%%%%%%%%%%%%%%%%%%%%%%%%%%%%%%%%%
%%%%%%%%%%%%%%%%%%%%  Section 3  %%%%%%%%%%%%%%%%%%%%%%%%%%
%%%%%%%%%%%%%%%%%%%%%%%%%%%%%%%%%%%%%%%%%%%%%%%%%%%%%%%%%%%

\section{A new model}

We now proceed to propose a new model, 
in which there is no fine-tuning of parameters. 
We introduce a global $U(1)_R$ symmetry 
but not the messenger gauge interaction. 
The $U(1)_R$ charges are assigned to each chiral 
superfield as shown in Table I. 
Under this $U(1)_R$ symmetry the first term in 
$W_{3-2}$ is forbidden. 
Instead, in the new model the superpotential 
is written as 
%%%%%%%%%%%%%%%%%%%%%%%%%%%%%%%%
\begin{equation}
\label{W}
      W = W_{SB} + W_{m} 
\end{equation}
%%%%%%%%%%%  W  %%%%%%%%%%%%%%%%
with 
%%%%%%%%%%%%%%%%%%%%%%%%%%%%%%%%%%%%%%%%%%%%%%%%%%%
\begin{eqnarray}
\label{WSB}
 W_{SB}   &=&  \frac {\lambda}{M^4} X_1 X_3 + 
                       \frac {\Lambda ^7}{X_3}, \\
 W_{m} &=&  \frac {h_1}{M^3} X_3 q \overline{q} + 
            \frac {h_2}{M^3} X_3 l \overline{l}\; , 
\end{eqnarray}
%%%%%%%%%%%  WSB  %%%%%%%%%%%%%%%%%%%%%%%%%%%%%%%%
where $\lambda $ and $h_i$'s are coupling constants 
taken to be of $O(1)$ and $M = M_{{\rm Planck}}$. 
The scale $\Lambda$ at which the $SU(3)$ gauge 
coupling becomes strong is expected to be sufficiently 
small compared with $M$. 
%%%%%%%%%%%  TABLE I  %%%%%%%%%%%%%%%%%
\vspace{5mm}
\begin{center}
\framebox[3cm] {\large \bf Table I}
\end{center}
\vspace{5mm}
%%%%%%%%%%%%%%%%%%%%%%%%%%%%%%%%%%%%%%%

The $D$-flat direction of the $SU(3) \times SU(2)$ 
is given by 
%%%%%%%%%%%%%%%%%%%%%%%%%%%%%%%%%%%%%%%%%
\begin{equation}
\begin{array}{rcl}
\label{D-flat}
  \langle Q \rangle & = & 
   \left(
     \begin{array}{cc}
            b & 0 \\
            0 & a \\
            0 & 0 
     \end{array}
   \right) , \qquad
  \langle \overline{U} \rangle = 
   \left(
     \begin{array}{c}
             b \ e^{i \theta} \\
             0  \\
             0  
     \end{array}
   \right) , \\
  \langle \overline{D} \rangle & = & 
   \left(
       \begin{array}{c}
             0  \\
             a \ e^{i \delta} \\ 
             0  
       \end{array}
   \right) , \qquad
  \langle L \rangle = 
           (- \sqrt{a^2 - b^2}\ e^{i \eta}, 0) .
\end{array}
\end{equation}
%%%%%%%%  D-flat  %%%%%%%%%%%%%%%%%%%%%%%%%%%%%%%%
Parameters $a$ and $b$ are real and positive 
with $a > b$. 
The symbols $\theta, \delta$, and $\eta$ 
stand for relative phases. 
Along the $D$-flat direction the scalar potential 
is expressed as 
%%%%%%%%%%%%%%%%%%%%%%%%%%%%%%%%%%%%%%%%%%%%%%%%%%%
\begin{equation}
\label{WKW}
    V = W_i \; (K^{-1})_{i \; \overline{j}} \; 
                              W_{\overline{j}}^* ,
\end{equation}
%%%%%%%%%%  WKW  %%%%%%%%%%%%%%%%%%%%%%%%%%%%%%%%%%
where $W_i \equiv \partial W/ \partial X_i$ and 
$K_{i \; \overline{j}} \equiv 
\partial^2 K / \partial X_i 
\partial X_{\overline{j}}^*$. 
The explicit form of $V$ is 
%%%%%%%%%%%%%%%%%%%%%%%%%%%%%%%%%%%%%%%%%%%%%%%
\begin{eqnarray}
\label{V1}
   V &=& \frac{2}{t} \, \left| 
            \frac{2 \lambda}{M^4} X_1 X_3
            - \frac{\Lambda^7}{X_3} 
            + \frac{h_1}{M^3}X_3 q \overline{q}
            + \frac{h_2}{M^3}X_3 l \overline{l} 
            \, \right|^2 
                                   \nonumber  \\
   {} & & + \;t^2 \, 
            \left( \frac{\lambda}{M^4} \right)^2 
            |X_3|^2 
          + 6B\,t \; \left| 
            \frac{\lambda}{M^4} X_1 
            - \frac{\Lambda^7}{X_3^2} 
            + \frac{h_1}{M^3} q \overline{q}
            + \frac{h_2}{M^3} l \overline{l} 
            \, \right|^2 
                                   \nonumber  \\
   {} & & + \left( \frac{h_1}{M^3} \right)^2 |X_3|^2 
              \left( |q|^2 + |\overline{q}|^2 \right)
            + \left( \frac{h_2}{M^3} \right)^2 |X_3|^2 
              \left( |l|^2 + |\overline{l}|^2 \right). 
\end{eqnarray}
%%%%%%%%%  V1  %%%%%%%%%%%%%%%%%%%%%%%%%%%%%%%%
Here we denote the scalar components of the superfield 
by the same letters as the superfield itself. 
It can be easily shown that at the minimum point 
of $V$ we have vanishing VEVs of $q$, $\overline{q}$ 
and $l$, $\overline{l}$. 
This is due to the fact that 
%%%%%%%%%%%%%%%%%%%%%%%%%%%%%%%%%%%%%%%%%
\begin{equation}
    \lambda \left( \frac{v}{M} \right)^4 
      \ll h_1 \left( \frac{v}{M} \right)^3, 
        \quad  h_2 \left( \frac{v}{M} \right)^3, 
\end{equation}
%%%%%%%%%%%%%%%%%%%%%%%%%%%%%%%%%%%%%%%%%
where the scale $v$ is redefined as 
%%%%%%%%%%%%%%%%%%%%%%%%%%%%%%%%%%%%%%
\begin{equation}
      v = \left( \frac{\Lambda^7 M^4}
           {\lambda} \right)^{1/11} 
\end{equation}
%%%%%%%%%%%%%%%%%%%%%%%%%%%%%%%%%%%%%%
in the new model and $\Lambda \ll v \ll M$. 
Further, since $V$ takes the minimum value along 
the direction $\cos (2\theta + 3\delta + \eta)=1$, 
we can put $\theta =\delta = \eta =0$ 
without loss of generality. 
The minimization of $V$ can be carried out 
numerically and results in 
%%%%%%%%%%%%%%%%%%%%%%%%%%%%%%%%%%%%%%%%%%%%%
\begin{equation}
  a = 1.072 \times v, \qquad \qquad b = 0.897 \times v. 
\end{equation}
%%%%%%%%%%%%%%%%%%%%%%%%%%%%%%%%%%%%%%%%%%%%%
In addition, we have the VEVs 
%%%%%%%%%%%%%%%%%%%%%%%%%%%%%%%%%%%%%%%%%%%%%%%%%%%%
\begin{eqnarray}
\label{VEV1}
  \langle X_1 \rangle & = & a^2 \sqrt{a^2-b^2} 
                        = 0.674 \times v^3,  \\
\label{VEV2}
  \langle X_3 \rangle & = & a^2 b^2 
                        = 0.924 \times v^4 
\end{eqnarray}
%%%%%%%%%%%  VEV1, VEV2  %%%%%%%%%%%%%%%%%%%%%%%%%%
and 
%%%%%%%%%%%%%%%%%%%%%%%%%%%%%%%%%%%%%%%%%%%%%%%%%%%%%%%%
\begin{eqnarray}
\label{VEVF1}
  \langle F_{X_1} \rangle & = & 1.412\times 
                       \lambda \frac {v^8}{M^4}, \\
\label{VEVF2}
  \langle F_{X_3} \rangle & = & -0.396 \times 
                       \lambda \frac {v^9}{M^4}. 
\end{eqnarray}
%%%%%%%%%%  VEVF1, VEVF2  %%%%%%%%%%%%%%%%%%%%%%%%%%%%%%
On the other hand, $\langle X_2 \rangle$ and 
$\langle F_{X_2} \rangle$ vanish. 
Then we obtain the ratios 
%%%%%%%%%%%%%%%%%%%%%%%%%%%%%%%%%%%%%%%%%%%
\begin{eqnarray}
\label{F/X1}
   \frac {\langle F_{X_1}\rangle }{\langle X_1 \rangle } 
     & = & 2.096 \times \lambda \frac{v^5}{M^4}, \\
\label{F/X2}
   \frac {\langle F_{X_3}\rangle }{\langle X_3 \rangle } 
     & = & -0.428 \times \lambda \frac{v^5}{M^4}. 
\end{eqnarray}
%%%%%%%%%%%%  F/X1, F/X2  %%%%%%%%%%%%%%%%%
It is now understandable that 
the factor $\lambda (v/M)^4$ corresponds 
to the very small parameter $\lambda_1$ 
in the previous paper. 
The new model naturally explains 
the smallness of $\lambda_1$.

Let us estimate gauge-mediated soft breaking 
masses of gauginos, squarks, and sleptons 
in the visible sector. 
Gaugino masses $m_{\lambda_a}$ are induced at 
the one-loop level \cite{456} and given by 
%%%%%%%%%%%%%%%%%%%%%%%%%%%%%%%%%%%%%%%%%%%%%%%%%
\begin{equation}
\label{gaugino}
 m_{\lambda_a} =  {g_a^2 \over 16 \pi^2}\;
    {\langle F_{X_3}\rangle \over \langle X_3 \rangle } ,
\end{equation}
%%%%%%%%%%%  gaugino  %%%%%%%%%%%%%%%%%%%%%%%%%%%
where $a$ denotes the index of $G_{st}$. 
Squark and slepton masses squared $\tilde{m}^2$ 
arise at the two-loop level \cite{456}
and are expressed as 
%%%%%%%%%%%%%%%%%%%%%%%%%%%%%%%%%%%%%%%%
\begin{equation}
\label{sparticle}
 \tilde{m}^2 =  \sum_a 2 \,C_F^a  
        \left( {g_a^2 \over 16 \pi^2} \right)^2 
            \left( {\langle F_{X_3} \rangle \over 
                    \langle X_3 \rangle }\right)^2 ,
\end{equation}
%%%%%%%%%%  sparticle  %%%%%%%%%%%%%%%%%%
where $C_F^a$ is $4/3$ for $SU(3)_c$ triplets, 
$3/4$ for $SU(2)_L$ doublets, 
and $3/5 Y^2$ for $U(1)_Y$. 
It is noteworthy that although the couplings 
$X_3 q \overline{q}$ and $X_3 l \overline{l}$ 
in $W_m$ are suppressed by inverse powers of 
the scale $M$, 
soft breaking masses in the visible sector 
do not have any suppression factors of $1/M$. 
{}From Eqs. (\ref{F/X1}) and (\ref{gaugino}), 
the relation 
%%%%%%%%%%%%%%%%%%%%%%%%%%%%%%%%%
\begin{equation}
\label{lamv}
    \lambda \; \frac{v^5}{M^4} \simeq 
                     10^{4 \sim 5} \,{\rm GeV} 
\end{equation}
%%%%%%%%  lamv  %%%%%%%%%%%%%%%%%
should be satisfied so as to obtain appropriate 
soft masses of order $10^{2 \sim 3}\,{\rm GeV}$. 
By setting $\lambda = O(1)$ and $M = M_{Planck}
= 10^{18.3}\,{\rm GeV}$, 
we are led to 
%%%%%%%%%%%%%%%%%%%%%%%%%%%%%%%%%%%%%%%%%
\begin{equation}
\label{vn}
    v   \sim 10^{-2.8} \times M 
            \simeq 10^{15.5} \,{\rm GeV}, 
\end{equation}
%%%%%%%%%%  vn  %%%%%%%%%%%%%%%%%%%%%%%%%
which is around the GUT scale. 
This implies that 
%%%%%%%%%%%%%%%%%%%%%%%%%%%%%%%%%
\begin{equation}
  \lambda \left( \frac{v}{M} \right)^4 \sim 10^{-11}, 
       \qquad \qquad \frac{v}{\Lambda} \sim 40. 
\end{equation}
%%%%%%%%%%%%%%%%%%%%%%%%%%%%%%%%%
Therefore, $\Lambda$ is of order $10^{14}\,{\rm GeV}$ 
and the perturbative picture is available.

In order to compare with the gauge-mediated 
contributions, 
we next estimate soft breaking masses of gauginos 
and sparticles transmitted by the 
gravitational interaction. 
Gaugino masses can arise from the operator 
%%%%%%%%%%%%%%%%%%%%%%%%%%%%%%%%%%%%%%%%%%%%%%%
\begin{equation}
    \int d^2 \theta \;{X_1 \over M^3} 
                             W_{\alpha} W^{\alpha} ,
\end{equation}
%%%%%%%%%%%%%%%%%%%%%%%%%%%%%%%%%%%%%%%%%%%%%%%
where $W^{\alpha}$ is the field strength 
superfield of the standard model gauge group. 
Thus the gravitational contribution to gaugino masses 
becomes 
%%%%%%%%%%%%%%%%%%%%%%%%%%%%%%%%%%%%%%%%%
\begin{equation}
\label{GM}
  m_{\lambda} \simeq {\langle F_{X_1}\rangle 
                              \over M^3} 
              \simeq \lambda {v^8 \over M^7}
              \sim 10^{-4} \,{\rm GeV}.
\end{equation}
%%%%%%%%%%%%%%%%%%%%%%%%%%%%%%%%%%%%%%%%%
Therefore, dominant contribution to gaugino masses 
comes from the standard model gauge interactions. 
On the other hand, soft masses squared of 
squarks and sleptons can arise from the operator 
%%%%%%%%%%%%%%%%%%%%%%%%%%%%%%%%%%%%%%%%%%%%%%%%%%%%
\begin{equation}
    \int d^4 \theta \;
     \; c \; {\phi^{\dag} \phi \over M^2} 
       \varphi^{\dag}_i \varphi_i 
\end{equation}
%%%%%%%%%%%%%%%%%%%%%%%%%%%%%%%%%%%%%%%%%%%%%%%%%%%
where $\phi$ is an elementary field of the 
DSB sector in Eq. (\ref{DSBfield}), 
$\varphi_i$ is the chiral superfield of 
the standard model 
and $c$ is a constant with $c \leq 1$ 
which represents the elementary field content 
of the composite fields $X_i$. 
Thus, the gravitational contribution to 
squark and slepton masses are given by 
%%%%%%%%%%%%%%%%%%%%%%%%%%%%%%%%%%%%%%%%%%%
\begin{equation}
\label{SQM}
  \tilde{m}^2  
           \sim  \left( {\langle F_{\phi}\rangle 
                                    \over M} \right)^2 
           \simeq \left( \lambda {v^6 \over M^5}\right)^2 
           \sim  \left( 10^{1.5}{\rm GeV}\right)^2,
\end{equation}
%%%%%%%%%%  SQM  %%%%%%%%%%%%%%%%%%%%%%%%%%
which are smaller than the gauge-mediated contribution 
by the factor $\sim 10^{-2}$.

In the present model mass spectra of other particles 
can be determined. 
Masses of vector-like fields 
$q, \overline{q}$ and $l, \overline{l}$ 
are of order $v^4/M^3 \sim 10^7 \,{\rm GeV}$. 
Three scalars, two pseudoscalars, and one fermion 
of moduli fields $X_i$ gain masses of order 
$v^5/M^4 \sim 10^4 \,{\rm GeV}$. 
The fermion component of $X_2$ remains massless 
because the global $U(1)$ symmetry in the 3\,-2 sector 
is unbroken down to low energies. 
Since $X_2$ does not appear in the superpotential 
of Eq. (\ref{W}), 
the massless fermion of $X_2$ couples very weakly 
only through the K\"ahler potential. 
Thus the global $U(1)$ and the massless fermion 
are harmless to our visible sector. 
The mass of the $R$-axion is $O(10^3 \,{\rm GeV})$, 
which is free from the constraint of supernova 
astrophysical bounds. 
From the condition on vanishing cosmological constant, 
the gravitino mass is 
%%%%%%%%%%%%%%%%%%%%%%%%%%%%%%%%%%%%%%%
\begin{equation}
\label{GRM}
   m_{3/2}  \simeq  {\sqrt{| \langle F_{\phi}\rangle |^2} 
                       \over \sqrt{3} M} 
            \simeq {\lambda \, v^6 \over \sqrt{3} M^5} 
            \sim    10^{1.5} \,{\rm GeV}.
\end{equation}
%%%%%%%%%%%  GRM  %%%%%%%%%%%%%%%%%%%%%%

It is worth noting that the soft supersymmetry 
breaking parameters are blind to flavor at the 
intermediate scale $\Lambda \sim 10^{14}\,{\rm GeV}$. 
As a consequence, enough degeneracy among 
squarks and sleptons is maintained at low energies 
and flavor changing neutral currents are naturally 
suppressed. 
As for the $SU(2)_L \times U(1)_Y$ breaking, 
it can be realized by the three-loop effects of 
the stop or by introducing another set of 
vector-like fields in the visible sector \cite{456}.

Since the present model incorporates the Planck 
scale in non-renormalizable interactions, 
we should use the scalar potential of $N=1$ 
supergravity for the correct discussion. 
However, even if we use the scalar potential of 
the supergravity, 
the above discussion remains unchanged.

%%%%%%%%%%%%%%%%%%%%%%%%%%%%%%%%%%%%%%%%%%%%%%%%%%%%%%%%%%%%%%
%%%%%%%%%%%%%%%%%%  Section 4  %%%%%%%%%%%%%%%%%%%%%%%%%%%%%%%
%%%%%%%%%%%%%%%%%%%%%%%%%%%%%%%%%%%%%%%%%%%%%%%%%%%%%%%%%%%%%%

\section{The $\mu$ problem}

In this section we explore a possible solution to 
the $\mu$ problem. 
As is well-known, 
phenomenological constraints on the Higgs potential 
including the vacuum stability condition 
require that the supersymmetric parameter $\mu$ and 
the soft supersymmetry breaking parameter $B$ 
satisfy the condition 
%%%%%%%%%%%%%%%%%%%%%%%%%%%%%%%%%%%%%%%%%%%%
\begin{equation}
\label{100}
     B \mu \sim \mu^2 \sim (100 \,{\rm GeV})^2. 
\end{equation}
%%%%%%%%%%%  100  %%%%%%%%%%%%%%%%%%%%%%%%%%
The minimal scheme of the messenger sector scenario 
has a difficulty in this regard. 
In the minimal messenger sector scenario, naively, 
the Higgs superpotential is written as 
%%%%%%%%%%%%%%%%%%%%%%%%%%%%%%%%%%%%%%%%%
\begin{equation}
    W_H = \lambda_H S H_u H_d\,, 
\end{equation}
%%%%%%%%%%%%%%%%%%%%%%%%%%%%%%%%%%%%%%%%%
where $S$ is a gauge singlet superfield with 
the non-vanishing $\langle S \rangle$ and 
$\langle F_S \rangle$. 
It follows that $\mu = \lambda_H \langle S \rangle$ 
and $B \mu = \lambda_H \langle F_S \rangle$. 
Then we have $B = \langle F_S \rangle/\langle S \rangle$
irrespective of the magnitude of $\lambda_H$. 
However, the ratio $\langle F_S \rangle/\langle S \rangle$ 
has to be of order $10^{4 \sim 5}\,{\rm GeV}$ to generate 
appropriate soft masses in the minimal scenario. 
Thus we have too large $B$ in disaccord with 
the above condition (\ref{100}).

Introducing an additional gauge singlet field $N$ with 
the non-zero VEV but with the vanishing 
$\langle F_N \rangle$ as well as Higgs fields 
$H_u$ and $H_d$ in the present model, 
we assume an alternative superpotential for 
Higgs fields. 
Namely, we add the Higgs terms 
%%%%%%%%%%%%%%%%%%%%%%%%%%%%%%%%%%%%%%%%%%%%%%%%%%%%
\begin{equation}
    W_H = \left( \frac{k_1}{M}N^2 + \frac{k_2}{M^4}
                      X_3 N \right) H_u H_d\, 
\end{equation}
%%%%%%%%%%%%%%%%%%%%%%%%%%%%%%%%%%%%%%%%%%%%%%%%%%%%
to the superpotential, 
where constants $k_1$ and $k_2$ are $O(1)$. 
Under the condition $\langle N \rangle \gg 
\langle X_3 \rangle /M^3$, 
the superpotential $W_H$ brings about 
%%%%%%%%%%%%%%%%%%%%%%%%%%%%%%%%%%%%%%%%%%%%%
\begin{eqnarray}
\label{mu}
    \mu   & \simeq & \frac{k_1}{M} 
                        \langle N \rangle ^2,  \\
\label{Bmu}
    B \mu & \simeq & \frac{k_2}{M^4} 
          \langle F_{X_3} \rangle \langle N \rangle. 
\end{eqnarray}
%%%%%%%%%%%  mu,  Bmu  %%%%%%%%%%%%%%%%%%%%%%
If we take the value 
$\langle N \rangle \sim 10^{10.5} \,{\rm GeV}$ 
as an ad hoc assumption here, 
then Eq. (\ref{mu}) yields $\mu = O(10^2\,{\rm GeV})$. 
Substituting the relation 
$|\langle F_{X_3} \rangle | \simeq \lambda v^9/M^4$ 
into Eq. (\ref{Bmu}), 
we obtain 
%%%%%%%%%%%%%%%%%%%%%%%%%%%%%%%%%%%%%%%%
\begin{equation}
   B \mu \simeq \frac{v^9}{M^8} \langle N \rangle. 
\end{equation}
%%%%%%%%%%%%%%%%%%%%%%%%%%%%%%%%%%%%%%%%
When we take Eq. (\ref{vn}) and 
$\langle N \rangle \simeq 10^{10.5} \,{\rm GeV}$ 
into account, 
the numerical value becomes 
%%%%%%%%%%%%%%%%%%%%%%%%%%%%%%%%%%
\begin{equation}
  B \mu \sim \left( 10^2 \,{\rm GeV} \right)^2. 
\end{equation}
%%%%%%%%%%%%%%%%%%%%%%%%%%%%%%%%%%
This result is in accord with the above conditions 
(\ref{100}) on $\mu$ and $B$. 
Further, the numerical value $\langle N \rangle \sim 
\sqrt{\mu M} \sim 10^{10.5} \,{\rm GeV}$ is also 
consistent with the condition $\langle N \rangle \gg 
\langle X_3 \rangle /M^3 (\sim 10^7 \,{\rm GeV})$. 
In this scenario it is essential that there exist 
not only the singlets $X_1$ and $X_3$ which 
communicate the supersymmetry breaking to the visible 
sector but also another singlet $N$ with 
the non-zero VEV but with the vanishing 
$\langle F_N \rangle$.

Although in the above we put 
$\langle N \rangle \sim 10^{10.5} \,{\rm GeV}$
by hand, 
it is not unreasonable. 
This scale of $\langle N \rangle$ is nearly equal to 
$\langle X_1 \rangle /M^2 \sim 10^{10.5}\,{\rm GeV}$. 
It is plausible that the field $N$ also corresponds 
to a composite operator. 
In the present model we have no such an operator 
fitting to the field $N$. 
However, when we take the gauge group larger than 
$SU(3) \times SU(2)$, 
the gauge singlet field fitting to the above 
$N$-field is possibly contained in the massless 
sector of the effective theory.

%%%%%%%%%%%%%%%%%%%%%%%%%%%%%%%%%%%%%%%%%%%%%%%%%%%%%%%%%%%%
%%%%%%%%%%%%%%%  Section 5  %%%%%%%%%%%%%%%%%%%%%%%%%%%%%%%%
%%%%%%%%%%%%%%%%%%%%%%%%%%%%%%%%%%%%%%%%%%%%%%%%%%%%%%%%%%%%

\section{Summary}

In this paper we proposed a new model in which 
suitable soft supersymmetry breaking masses are 
dynamically generated without the messenger sector. 
The model proposed here is free from the fine-tuning 
of coupling constants. 
The model is a simple extension of the 3\,-2 model. 
The dynamical supersymmetry breaking sector 
contains extra vector-like quarks and leptons 
which have the gauge quantum number 
of the standard model. 
Non-renormalizable interactions play a crucial 
role of transmitting supersymmetry breaking to 
the visible sector through the gauge interaction. 
The non-renormalizable interactions are 
characterized by the couplings of $O(1)$ in units 
of the reduced Planck scale $M_{\rm Planck}$. 
Hierarchy of various mass scales arises from the 
non-renormalizable terms. 
The scale of the DSB is of order 
$10^{14} \,{\rm GeV}$. 
This model conserves the color symmetry. 
Although the supersymmetry breaking is transmitted 
by both gravity and gauge interactions, 
gaugino, squark, and slepton soft masses mainly 
come from the gauge-mediated contributions. 
Consequently, we obtain enough degeneracy between 
sfermions of the same gauge quantum numbers 
to ensure adequate suppression of FCNC. 
The gravitino mass is of order $10\,{\rm GeV}$.

We discussed a possible solution to the $\mu$ 
problem in the present framework. 
It is necessary for us to introduce an additional 
gauge singlet $N$ with the non-zero VEV but with 
the vanishing $\langle F_N \rangle$. 
When we take a large gauge group instead of 
$SU(3) \times SU(2)$, 
the massless sector of the effective theory 
may contain such a singlet. 
The present model will provide a useful guide 
to constructing the phenomenological viable models 
of gauge-mediated supersymmetry breaking.

\vspace{5cm}
{\bf Acknowledgements}

The authors thank Prof. Kitakado for careful 
reading of the manuscript. 
They also thank C. Cs\'aki and W. Skiba 
for a valuable comment. 
This work is supported in part by the Grant-in-Aid 
for Scientific Research, Ministry of Education, 
Science and Culture, Japan (No. 08640366).

%%%%%%%%%%%%%%%%%%%%%%%%%%%%%%%%%%%%%%%%%%%%%%%%%%%%%%%%%%%%%%%%
%%%%%%%%%%%%  References  %%%%%%%%%%%%%%%%%%%%%%%%%%%%%%%%%%%%%%%
%%%%%%%%%%%%%%%%%%%%%%%%%%%%%%%%%%%%%%%%%%%%%%%%%%%%%%%%%%%%%%%%%

%%%%%%%%%%%%%%%%%%%%%%%%%%%%%%%%%%%%%%%%%%%%%%%%%%%%%
%%%%%%%%%%%  TABLE  %%%%%%%%%%%%%%%%%%%%%%%%%%%%%%%%%
%%%%%%%%%%%%%%%%%%%%%%%%%%%%%%%%%%%%%%%%%%%%%%%%%%%%%
\newpage

\begin{center}
{\Large\bf Table Captions}
\end{center}

\begin{flushleft}
{\bf Table I} 
\end{flushleft}

\noindent
The $U(1)_R$-charges of matter superfields and 
their $SU(3) \times SU(2)$ gauge invariant operators. 
The parameter $r$ is arbitrary other than $-5/3$.

\vspace{4cm}

\begin{center}

{\bf Table I} \\

\vspace{6mm}

\begin{tabular}{|c|c|}  \hline
\vphantom{\bigg(}
   superfields          &    $U(1)_R$-charges  \\
\hline
\vphantom{\bigg(}
        $Q\;$            &       $r$        \\
 ${\overline U}$         &     $-6-4r$      \\
 ${\overline D}$         &      $4+2r$      \\
        $L\;$            &       $-3r$      \\
\hline \hline
\vphantom{\bigg(}
 $X_1 = Q{\overline D}L\;$          &    $4$      \\
 $X_2 = Q{\overline U}L$            &  $-6-6r$    \\
 $\ \ \;X_3 = \det Q{\overline Q}$  &   $-2$      \\
 $q{\overline q},\ l{\overline l}$  &    $4$      \\
\hline
\end{tabular}
\end{center}

\end{document}